\documentclass[11pt]{article}
\usepackage{amsmath}
\usepackage{amssymb}
\usepackage{amsmath,amsthm,epsfig,euscript,array,cite,cancel}
\usepackage{mathtools}

\newcommand{\be}{\begin{equation}}
\newcommand{\bea}{\begin{eqnarray}}
\newcommand{\ee}{\end{equation}}
\newcommand{\eea}{\end{eqnarray}}

\begin{document}

\newcommand{\ds}{\displaystyle}

\topmargin -1cm \oddsidemargin=0.25cm\evensidemargin=0.25cm
\setcounter{page}0
\renewcommand{\thefootnote}{\fnsymbol{footnote}}
\begin{titlepage}
\begin{flushright}
CERN-PH-TH/2014-175 \\
MPP-2014-331
\end{flushright}
\vskip .7in
\begin{center}
{\LARGE \bf Higher Spins in Hyper-Superspace \\
}
\vskip .6in {\Large Ioannis Florakis$^{a,b}$\footnote{e-mail: {\tt ioannis.florakis@cern.ch}},
Dmitri Sorokin$^c$\footnote{e-mail: {\tt  dmitri.sorokin@pd.infn.it }} and
 Mirian Tsulaia$^d$\footnote{e-mail: {\tt  mirian.tsulaia@canberra.edu.au}  }}
 \vskip .4in {$^a$ \it Department of Physics, CERN - Theory Division, CH-1211 Geneva 23, Switzerland} \\
\vskip .2in {$^b$ \it Max-Planck-Institut f\"ur Physik, Werner-Heisenberg-Institut, 80805 M\"unchen, Germany} \\
\vskip .2in {$^c$ \it INFN, Sezione di Padova, via F. Marzolo 8, 35131 Padova, Italia} \\
\vskip .2in { $^d$ \it Faculty of  Education, Science, Technology and Mathematics, University of
Canberra, Bruce ACT 2617, Australia}\\
\vskip .8in

\begin{abstract}

We extend the results of arXiv:1401.1645 on the generalized conformal $Sp(2n)$--structure of infinite multiplets of higher-- spin fields, formulated in spaces with extra tensorial directions (hyperspaces), to the description of $OSp(1|2n)$--invariant infinite--dimensional higher--spin supermultiplets formulated in terms of scalar superfields on flat hyper--superspaces and on $OSp(1|n)$ supergroup manifolds. We find generalized superconformal transformations relating the superfields and their equations of motion in flat hyper--superspace with those on the $OSp(1|n)$ supermanifold. We then use these transformations to relate the two--, three-- and four--point correlation functions of the scalar superfields on flat hyperspace, derived by requiring the $OSp(1|2n)$ invariance of the correlators, to correlation functions on the $OSp(1|n)$ group manifold. As a byproduct, for the simplest particular case of a conventional ${\mathcal N}=1$, $D=3$ superconformal theory of scalar superfields, we also derive correlation functions of component fields of the scalar supermultiplet including those of auxiliary fields.

\end{abstract}

\end{center}

\vfill

\end{titlepage}

\renewcommand{\thefootnote}{\arabic{footnote}}
\setcounter{footnote}0

\section{Introduction}
In \cite{Florakis:2014kfa} we have studied some aspects of the description of infinite sets of integer and half--integer massless
 higher-- spin fields in flat and anti--de--Sitter (AdS) spaces in terms of scalar and spinor `hyperfields' propagating in hyperspaces. In addition to a conventional space--time as a subspace, hyperspaces are endowed with extra tensorial coordinates encoding  the spin degrees of freedom of conventional space--time fields. This formulation, which was originally put forward by Fronsdal as an alternative to the Kaluza--Klein theory \cite{Fronsdal:1985pd}, has been extensively developed by several authors \cite{Bandos:1999qf,Vasiliev:2001zy,Vasiliev:2001dc,Didenko:2003aa,Plyushchay:2003gv,Plyushchay:2003tj,Gelfond:2003vh,Vasiliev:2003jc,Bandos:2004nn,Bandos:2005mb,Vasiliev:2007yc,Ivanov:2007vx,West:2007mh,Gelfond:2008ur,Gelfond:2008td,Gelfond:2010xs,Gelfond:2010pm,Bandos:2011wi,Fedoruk:2012ka,Gelfond:2013lba}.

The theories on tensorially extended (super)spaces, which we will henceforth refer to as hyper-(super)spaces,
offer many interesting and challenging problems regarding higher--spin fields,  one of them
being the further development and study of  generalized (super)conformal theories on these spaces. This motivated our recent work \cite{Florakis:2014kfa} in which, using generalized conformal transformations, we established an explicit relation between the equations of motion of hyperfields on flat hyperspace and on $Sp(n)$ group--manifolds, the latter being tensorial generalizations\footnote{Here $Sp(n)$ stands for the real non--compact form $Sp(n,\mathbb R)$ of $Sp(n,\mathbb C)$, where $\mathbb R$  will be omitted for brevity.} of AdS spaces.
This relation was then employed in order to explicitly derive the $Sp(2n)$--invariant two--, three-- and four--point correlation functions for
 fields on $Sp(n)$ group manifolds, from the known $Sp(2n)$--invariant correlation functions on flat hyperspaces,
thus,  generalizing the  results obtained   in  \cite{Vasiliev:2001dc,Vasiliev:2003jc,Didenko:2012tv}.

In this paper we further extend the results of \cite{Florakis:2014kfa} to the description of supersymmetric systems of higher-- spin fields in hyper--superspaces, which were  previously studied \emph{e.g.} in \cite{Bandos:1999qf,Bandos:1999pq,Vasiliev:2001zy,Didenko:2003aa,Plyushchay:2003gv,Plyushchay:2003tj,Bandos:2004nn,Ivanov:2007vx,Bandos:2011wi}. In particular, by means of a generalized superconformal transformation, we
 establish an explicit relation between the superfield equations of motion \cite{Bandos:2004nn} on flat hyper--superspace and on an $OSp(1|n)$ supergroup manifold.
Furthermore, the explicit solution of the generalized superconformal Ward identities allows us to derive the $OSp(1|2n)$--invariant two--, three-- and four--point superfield correlation functions on flat hyper--superspace and, consequently, using the generalized superconformal transformations, we obtain the corresponding correlation functions on the $OSp(1|n)$ group manifolds. Our results, therefore, generalize the superfield description and computation of superfield correlators in conventional superconformal field theories, considered \emph{e.g.} in \cite{Park:1997bq,Park:1999cw,Kuzenko:1999pi,Dolan:2000ut}, to superconformal higher--spin theories. A byproduct of our analysis is the derivation of  correlation functions involving the component fields of the scalar supermultiplet, including the auxiliary fields, for the simple special case of a three-dimensional  ${\mathcal N}=1$ superconformal theory of scalar superfields.

As in the case of the ${\mathcal N}=1$, $D=3$ superconformal theory, the fact that 3-- and 4--point correlation functions are non-zero for hyperfields of an anomalous conformal weight may indicate the existence of interacting conformal higher--spin fields which involve higher orders of their field strengths.

It should be noted that in the literature \cite{Curtright:1979uz,Vasiliev:1980as,Bellon:1986ki,Fradkin:1987ah,Bergshoeff:1988jm,Konstein:1989ij,
Kuzenko:1993jp,Kuzenko:1993jq,Kuzenko:1994dm,Buchbinder:1995ez,Gates:1996my,Gates:1996xs,Sezgin:1998gg,
Alkalaev:2002rq,Engquist:2002gy,Sezgin:2012ag,Zinoviev:2007js,Fotopoulos:2008ka,Gates:2013rka,Gates:2013ska, Candu:2014yva} various supersymmetric higher--spin systems have been considered in either irreducible
or reducible representations of the Poincar\'e and AdS groups (see \emph{e.g.} \cite{Francia:2002pt,Campoleoni:2012th}
 for a discussion of reducible higher-- spin multiplets in the ``metric--like" approach).
As we will see, the systems of integer and half--integer higher-- spin fields considered in  \cite{Bandos:1999qf,Vasiliev:2001zy,Didenko:2003aa,Plyushchay:2003gv,Plyushchay:2003tj,Bandos:2004nn,Ivanov:2007vx,Bandos:2011wi}
and in this paper form \emph{irreducible} infinite--dimensional supermultiplets of space--time supersymmetry. These supersymmetric
 higher--spin systems are therefore different from finite--dimensional higher--spin supermultiplets considered  in
\cite{Curtright:1979uz,Vasiliev:1980as,Bellon:1986ki,Fradkin:1987ah,Bergshoeff:1988jm,Konstein:1989ij,
Kuzenko:1993jp,Kuzenko:1993jq,Kuzenko:1994dm,Buchbinder:1995ez,Gates:1996my,Gates:1996xs,Sezgin:1998gg,
Alkalaev:2002rq,Engquist:2002gy,Sezgin:2012ag,Zinoviev:2007js,Fotopoulos:2008ka,Gates:2013rka,Gates:2013ska, Candu:2014yva}. We will provide the algebraic reasoning for this in Section 2.4.

The paper is organized as follows. Section \ref{s2} begins with a review of some basic known results about hyper-superspaces.
 We describe in detail the generalized superconformal algebra, the realization of the
generalized superconformal group $OSp(1|2n)$ on hyper-superspace and the precise connection between generalized and conventional conformal weights for scalar superfields and their components in various dimensions. Finally, we demonstrate how an infinite--dimensional $\mathcal N=1$ supersymmetry multiplet is formed by the component fields of the hyper--superfield in the case of four--dimensional flat space--time.

In Section \ref{s3} we provide a description of the geometric structure of $OSp(1|n)$ manifolds.
These manifolds exhibit the property of generalized superconformal flatness (or $GL$--flatness) observed earlier in
\cite{Plyushchay:2003gv,Plyushchay:2003tj}, which is similar to the superconformal flatness property of certain conventional AdS superspaces and superspheres \cite{Bandos:2002nn,Kuzenko:2011rd,Kuzenko:2012bc,Samsonov:2014pya,Kuzenko:2014yia}.
We then consider the relation between the $OSp(1|2n)$--invariant field equations for scalar superfields on flat hyper--superspace and those on the $OSp(1|n)$ group manifold derived in \cite{Bandos:2004nn}.
We show that, similarly to the non-supersymmetric case
\cite{Florakis:2014kfa}, the supersymmetric field equations on flat hyper-superspace and on $OSp(1|n)$ group manifolds  are related to each other
via a generalized superconformal transformation of the scalar hyper-superfield and its derivatives.

In Section \ref{s4}, as a preparation for the computation of correlation functions
on flat hyper-superspace and on $OSp(1|n)$ supergroup manifolds,
we consider the simplest example of an $OSp(1|4)$--invariant superconformal theory of a conventional $\mathcal N=1$, $D=3$ massless scalar superfield. Even though higher--spin fields are absent in this case, it is a simple setup in which one can illustrate the salient features of our approach. To this end, we present the $OSp(1|4)$--invariant two--, three-- and four--point correlation  functions of scalar superfields, as well as the correlators of the component fields of the scalar supermultiplet, including those of  auxiliary fields.

Finally, in  Section \ref{s5} we use the requirement of
$OSp(1|2n)$ invariance to derive the expressions for two--, three-- and four--point correlation functions
of the scalar hyper--superfields.
Again, in a complete analogy with the non--supersymmetric systems \cite{Florakis:2014kfa},
the correlation functions on flat hyper-superspaces and $OSp(1|n)$ supergroup
manifolds are related via generalized superconformal Weyl rescaling. Thus, our basic result is that the  $GL$--flatness is a key property of $Sp(n)$ and $OSp(1|n)$ manifolds that renders them amenable to the same type of analysis as for the case of flat hyper (super) spaces.

We conclude with a discussion on open problems and perspectives for further development of the hyperspace formulation of higher--spin fields.

\section{Scalar superfields in flat hyper--superspace, equations of motion and correlators} \label{s2} \setcounter{equation}0
\subsection{Flat hyper--superspace and its symmetries}\label{fss}
The flat hyper--superspace (see \emph{e.g.} \cite{Bandos:1999qf,Vasiliev:2001zy,Bandos:2004nn}) is parametrized by $\frac{n(n+1)}{2}$ bosonic matrix coordinates $X^{\mu\nu}=X^{\nu\mu}$ and $n$ real Grassmann--odd `spinor' coordinates $\theta^\mu$ ($\mu=1,\cdots,n$). We call $\theta^\mu$ `spinors', since they are indeed so from the perspective of  conventional space--time, which is a subspace of hyperspace.

For instance, when $n=4$, we can decompose the ten bosonic coordinates $X^{\mu\nu}$ using the Majorana (real) representation of the gamma--matrices of a $D=4$ space--time as follows
\be\label{X}
X^{\mu\nu}=X^{\nu\mu}=\frac 12\,x^m\,(\gamma_m)^{\mu\nu}+\frac 14
\,y^{mn}\,(\gamma_{mn})^{\mu\nu}\ , \qquad
\mu,\nu=1,2,3,4\,,\qquad m,n=0,1,2,3,
\ee
where $(\gamma_m)^{\mu\nu}=(\gamma_m)^{\nu\mu} \equiv C^{\mu\tau}(\gamma_m)_\tau {}^\nu$, $(\gamma_{mn})^{\mu\nu}=(\gamma_{mn})^{\nu\mu} \equiv C^{\mu\tau}(\gamma_{mn})_\tau {}^\nu$, with $C^T=-C$ being the charge conjugation matrix and the gamma--matrices $(\gamma_m)_{\mu}{}^\nu$ satisfy the Clifford algebra  $\{\gamma_m,\gamma_n\}=2\eta_{mn}$.
The space--time metric signature is chosen to be mostly plus $(-,+,\cdots,+)$.

The four coordinates $x^m$ parametrize the conventional flat space--time which is extended to flat hyperspace by adding six extra dimensions, parametrized by $y^{mn}=-y^{nm}$. This bosonic hyperspace is then further extended to the hyper--superspace by adding four Grassmann--odd directions parametrized by $\theta^\mu$, which transform in the spinor representation of the $D=4$ Lorentz group $SO(1,3)$.

The supersymmetry variation of the coordinates
\begin{equation}\label{susyv}
\delta\theta^\mu=\epsilon^\mu,\qquad \delta X^{\mu\nu}=-i\epsilon^{(\mu}\theta^{\nu)}
\,,
\end{equation}
leaves invariant the Volkov--Akulov--type one--form
\be\label{VA}
\Pi^{\mu\nu}=dX^{\mu\nu}+i\theta^{(\mu}d\theta^{\nu)}\,.
\ee
The round brackets denote symmetrization of  indices with the standard normalization
\be\label{sim}
Y^{(\mu_1\cdots\mu_k)}=\frac 1{k!}(Y^{\mu_1\cdots\mu_k}+\,\,\rm{all~permutations~ of~indices}\,)\,.
\ee
The supersymmetry transformations form a generalized super--translation algebra
\be\label{ST}
\{Q_\mu,Q_{\nu}\}=2P_{\mu\nu},\qquad [Q_\mu,P_{\nu\rho}]=0\,, \qquad [P_{\mu\nu},P_{\rho\lambda}]=0\,,
\ee
with $P_{\mu\nu}$ generating translations along $X^{\mu\nu}$. Namely, $\delta X^{\mu\nu}=ia^{\rho\lambda}P_{\rho\lambda}\cdot X^{\mu\nu}=a^{\mu\nu}$, with $a^{\mu\nu}$ being constant parameters.

The realization of $P_{\mu\nu}$ and $Q_\mu$ as differential operators is given by
\be\label{PQ}
P_{\mu\nu}=-i\frac{\partial}{\partial X^{\mu\nu}}\equiv-i\partial_{\mu\nu}\,,\qquad Q_{\mu}=\partial_\mu-i\theta^\nu\partial_{\nu\mu}\,,\qquad \partial_\mu\equiv \frac\partial{\partial\theta^\mu}\,,
\ee
where, by definition,
\be\label{pX}
\partial_{\mu\nu}\,X^{\rho\lambda}=\delta^{(\rho}_\mu\,\delta^{\lambda)}_\nu\,.
\ee
Furthermore, in the case $n=4$, $D=4$, the partial derivative associated with \eqref{X} takes the form
\be\label{pX4}
\partial_{\mu\nu}=\frac 12 (\gamma^m)_{\mu\nu}\frac{\partial}{\partial x^m}+\frac 12 (\gamma^{mn})_{\mu\nu}\frac{\partial}{\partial y^{mn}}\,.
\ee

The algebra \eqref{ST} is invariant under  rigid $GL(n)$ transformations
\be\label{GL}
Q'_{\mu}=g_{\mu}{}^{\nu}\,Q_\nu\,,\qquad P'_{\mu\nu}=g_{\mu}{}^\rho\,g_\nu{}^\lambda\,P_{\rho\lambda},
\ee
generated by
\be\label{L}
L_{\mu}{}^{\nu}=-{2i}(X^{\nu\rho}+\frac i2 \theta^\nu\theta^\rho)\partial_{\rho\mu}-i\theta^\nu\,Q_\mu\,,
\ee
which act on $P_{\mu\nu}$ and $Q_\mu$ as
\be\label{LPLQ}
[P_{\mu \nu},L_\lambda{}^\rho]=-i(\delta_\mu^\rho P_{\nu \lambda}+ \delta^\rho_\nu P_{\mu \lambda} )\,,\qquad [Q_{\mu},L_\nu{}^{\rho}]=-i\delta_{\mu}^{\rho}\,Q_\nu \,,
\ee
and close into the $gl(n)$ algebra
\be\label{gl}
[L_{\nu}{}^{\mu},L_{\lambda}{}^{\rho}]= i(\delta^{\mu}_\lambda\,L_{\nu}{}^{\rho}-\delta^{\rho}_\nu\,L_{\lambda}{}^{\mu})\,.
\ee
The algebra \eqref{ST}, \eqref{LPLQ} and \eqref{gl} is the hyperspace counterpart of the conventional super--Poincar\'e algebra enlarged by dilatations. That this is so can be most easily seen by considering \emph{e.g.}  \hbox{$n=2$} (i.e. $\mu=1,2$), in which case this algebra is recognized as the \hbox{$D=3$} super--Poincar\'e algebra with \hbox{$L_\mu{}^\nu-\frac 12 \delta_\mu^{\nu}\,L_\rho{}^\rho=M_m(\gamma^m)_\mu{}^{\nu}$} generating the $SL(2,R)\sim SO(1,2)$ Lorentz rotations (note that $m=0,1,2$)
and ${\mathbf D}=\frac 12 L_{\rho}{}^\rho$ being the dilatation generator. Note that the factor $\frac 12$ in the definition of the dilatation generator is required in order to have the canonical scaling of the momentum generator $P_{\mu\nu}$ with weight 1 and the supercharge $Q_\mu$ with weight $\frac 12$, as  follows from eq. \eqref{LPLQ}.

This algebra may be further extended to the $OSp(1|2n)$ algebra, generating generalized superconformal transformations of the flat hyper--superspace, by adding the additional set of supersymmetry generators
\be\label{S}
S^\mu=-(X^{\mu\nu}+\frac i2\theta^\mu\theta^\nu)Q_\nu\,,
\ee
together with the generalized conformal boosts
\be\label{K}
K^{\mu\nu}=i(X^{\mu\rho}+\frac i2\theta^\mu\theta^\rho)(X^{\nu\lambda}+\frac i2\theta^\nu\theta^\lambda)\partial_{\rho\lambda}-i\theta^{(\mu}S^{\nu)}\,.
\ee
The generators $S^\mu$ and $K^{\mu\nu}$ form a superalgebra similar to \eqref{ST}
\be\label{STd}
\{S^\mu,S^{\nu}\}=-2K^{\mu\nu},\qquad [S^\mu,K^{\nu\rho}]=0\,, \qquad [K^{\mu\nu},K^{\rho\lambda}]=0\,,
\ee
while the non--zero (anti)commutators of $S^\mu$ and $K^{\mu\nu}$ with $Q_\mu$, $P_{\mu\nu}$ and $L_\mu{}^\nu$ read
\be\label{QSP}
\{Q_\mu,S^\nu\}=-L_\mu{}^\nu\,,\quad [S^\mu,P_{\nu\rho}]=i\delta^\mu_{(\nu}\,Q_{\rho)}, \quad [Q_\mu,K^{\nu\rho}]=-i\delta_\mu^{(\nu}\,S^{\rho)}\,,\quad [S^\mu,L_\nu{}^\rho]=i\delta^\mu_\nu\,S^\rho\,.
\ee

\subsection{Generalized superconformal algebra $OSp(1|2n)$}
We now collect together all the non--zero (anti)commutation relations among the generators of the $OSp(1|2n)$ algebra
\bea\label{osp12n}
&\{Q_\mu,Q_{\nu}\}=2P_{\mu\nu},\qquad [Q_\mu,P_{\nu\rho}]=0\,, \qquad [P_{\mu\nu},P_{\rho\lambda}]=0\,,&\nonumber\\
&&\nonumber\\
&\{S^\mu,S^{\nu}\}=-2K^{\mu\nu},\qquad [S^\mu,K^{\nu\rho}]=0\,, \qquad [K^{\mu\nu},K^{\rho\lambda}]=0\,,&\nonumber\\
&&\nonumber\\
&\{Q_\mu,S^\nu\}=-L_\mu{}^\nu\,,\qquad [S^\mu,P_{\nu\rho}]=i\delta^\mu_{(\nu}\,Q_{\rho)}, \qquad [Q_\mu,K^{\nu\rho}]=-i\delta_\mu^{(\nu}\,S^{\rho)}\,,&\\
&&\nonumber\\
&[P_{\mu \nu},L_\lambda{}^\rho]=-i(\delta_\mu^\rho P_{\nu \lambda}+ \delta^\rho_\nu P_{\mu \lambda} )\,,\qquad [Q_{\mu},L_\nu{}^{\rho}]=-i\delta_{\mu}^{\rho}\,Q_\nu\,\qquad [S^\mu,L_\nu{}^\rho]=i\delta^\mu_\nu\,S^\rho\,,&\nonumber\\
&&\nonumber\\
&[L_{\nu}{}^{\mu},L_{\lambda}{}^{\rho}]= i(\delta^{\mu}_\lambda\,L_{\nu}{}^{\rho}-\delta^{\rho}_\nu\,L_{\lambda}{}^{\mu})\,, &\nonumber\\
&&\nonumber\\
&[K^{\mu \nu},L_\lambda{}^\rho]=i  ( \delta^\mu_\lambda  K^{\nu \rho}   + \delta^\nu_\lambda  K^{\mu \rho}  )\,,\qquad [P_{\mu \nu},K^{\lambda \rho}]=\frac{i}{4} (  \delta^\rho_\mu L_\nu{}^\lambda +  \delta^\rho_\nu L_\mu{}^\lambda
+  \delta^\lambda_\mu L_\nu{}^\rho +  \delta^\lambda_\nu L_\mu{}^\rho)  \,. &\nonumber
\eea
Let us note that in the case $n=4$, in which the physical space--time is four--dimensional (see eq. \eqref{X}) the generalized superconformal group $OSp(1|8)$ contains the $D=4$ conformal symmetry group $SO(2,4)\sim SU(2,2)$ as a subgroup, but  not the superconformal group $SU(2,2|1)$. The reason being that, although $OSp(1|8)$ and $SU(2,2|1)$ contain the same number of  (eight) generators, the anticommutators of the former close on the generators of the whole $Sp(8)$, while those of the latter only close on an $U(2,2)$ subgroup of $Sp(8)$, and the same supersymmetry generators cannot satisfy the different anti--commutation relations simultaneously. In fact, the minimal $OSp$--supergroup containing $SU(2,2|1)$ as a subgroup is $OSp(2|8)$.

\subsection{Scalar superfields and their $OSp(1|2n)$--invariant equations of motion}

Let us now consider a superfield $\Phi(X,\theta)$ transforming as a scalar under the super--translations given in eq.  \eqref{PQ}
\be\label{dPhi}
\delta \Phi=-(\epsilon^\alpha Q_\alpha\,+ia^{\mu\nu}P_{\mu\nu})\,\Phi\,.
\ee
To construct equations of motion for $\Phi(X,\theta)$ which are invariant under \eqref{dPhi} and comprise the equations of motion of an infinite tower of integer and half--integer higher-- spin fields with respect to conventional space--time, we introduce the spinorial covariant derivatives
\be\label{D}
D_\mu=\partial_\mu+i\theta^\nu\partial_{\nu\mu}\,, \qquad \{D_\mu,D_\nu\}=2i\partial_{\mu\nu}\,,
\ee
which (anti)commute with $Q_\mu$ and $P_{\mu\nu}$.

The $\Phi$--superfield equations then take  the  form \cite{Bandos:2004nn}
\be\label{DDPhi}
D_{[\mu}D_{\nu]}\Phi=0\,,
\ee
where the brackets denote the anti--symmetrization of indices with unit overall strength similarly to \eqref{sim}.
As was shown in \cite{Bandos:2004nn}, these superfield equations imply that all components of $\Phi(X,\theta)$ except for the first and the second one in the $\theta^\mu$--expansion of  $\Phi(X,\theta)$  should vanish
\be\label{bftheta}
\Phi(X,\theta)=b(X)+i\theta^\mu\,f_\mu(X)+i\theta^{\mu} \theta^\nu A_{\mu\nu}+\cdots\,,
\ee
(i.e. $A_{\mu_1\ldots \nu_k}=0$ for $k>1$) while the scalar and spinor fields $b(X)$ and $f_\mu(X)$ satisfy the equations first derived in \cite{Vasiliev:2001zy}
\begin{equation}\label{BF}
(\partial_{\mu \nu} \partial_{ \rho \lambda}- \partial_{\mu \rho} \partial_{\nu \lambda}) b(X)=0\,,
\end{equation}
\begin{equation}\label{FF}
\partial_{\mu \nu}f_\rho(X) -\partial_{\mu \rho}f_\nu(X)=0\,.
\end{equation}
For $n$=4, 8 and 16 these equations encode the Bianchi identity and equations of motion for the curvatures of infinite towers of conformally invariant, massless higher--spin fields in 4--, 6-- and 10--dimensional flat space--time, respectively (see \cite{Vasiliev:2001zy,Bandos:2005mb}).

The superfield equations \eqref{DDPhi} are invariant under the generalized superconformal $OSp(1|2n)$ symmetry, provided that $\Phi(X,\theta)$ transforms
as a scalar superfield with the `canonical' generalized scaling weight $\frac 12$, \emph{i.e.}
\bea\label{deltaPhi}
\delta\Phi &=&-(\epsilon^\mu\,Q_\mu+\xi_\mu\, S^\mu+ia^{\mu\nu}\,P_{\mu\nu}+ik_{\mu\nu}\,K^{\mu\nu}+ig_\mu{}^\nu\,L_{\nu}{}^\mu)\,\Phi\nonumber\\
&&-\frac 12\, \left(g_\mu{}^\mu-
 k_{\mu \nu} (X^{\mu \nu}+\frac i2 \theta^\mu\theta^\nu)+\xi_\mu\,\theta^\mu\right)\,\Phi\,,
\eea
where the factor $\frac 12$ in the second line is the generalized conformal weight and $\epsilon^\mu$, $\xi_\mu$, $a^{\mu\nu}$, $k_{\mu\nu}$ and $g_\mu{}^\nu$ are the rigid parameters of the $OSp(1|2n)$ transformations.

Scalar superfields with  anomalous generalized conformal dimension $\Delta$ transform under $OSp(1|2n)$  as
\bea\label{deltaPhiD}
\delta\Phi &=&-(\epsilon^\mu\,Q_\mu+\xi_\mu\, S^\mu+ia^{\mu\nu}\,P_{\mu\nu}+ik_{\mu\nu}\,K^{\mu\nu}+ig_\mu{}^\nu\,L_{\nu}{}^\mu)\,\Phi\nonumber\\
&&-\Delta\, \left(g_\mu{}^\mu-
 k_{\mu \nu} (X^{\mu \nu}+\frac i2 \theta^\mu\theta^\nu)+\xi_\mu\,\theta^\mu\right)\,\Phi\,.
\eea
It is instructive to demonstrate how the generalized conformal dimension $\Delta$, which is defined to be the same for all values of $n$ in $OSp(1|2n)$, is related to the conventional conformal weight of scalar superfields in various space--time dimensions. As we have already mentioned in Section \ref{fss}, the dilatation operator should be identified with $\mathbf D=\frac 12 L_{\mu}{}^\mu$. Therefore, considering a $GL(n)$ transformation \eqref{deltaPhiD} with parameter $g_{\mu}{}^{\nu}$
$$
\delta \Phi =-ig_\mu{}^\nu\,L_{\nu}{}^\mu\Phi,
$$
the part of the transformation corresponding to the dilatation reads
\be\label{Dt}
\delta_{\mathbf D} \Phi=-\frac in g_\mu{}^{\mu}\,L_{\nu}{}^\nu\Phi=-\frac {2i}n g_{\mu}{}^\mu {\mathbf D}\Phi=-i \tilde g {\mathbf D}\Phi\,,
\ee
where $\tilde g=\frac 2n g_{\mu}{}^\mu$ is the genuine dilatation parameter. From \eqref{deltaPhiD} it then follows that the conventional conformal weight $\Delta_D$ of the scalar superfield is related to the generalized one $\Delta$ via
\be\label{DD}
\Delta_D=\frac n2 \Delta\,.
\ee
In the $n=2$ case  corresponding to the ${\mathcal N}=1$, $D=3$ scalar superfield theory the two conformal dimensions coincide, whereas in the case $n=4$ describing conformal higher-- spin fields in $D=4$ one finds $\Delta_4=2\Delta$.
 Relation \eqref{DD} indeed provides the correct conformal dimensions of scalar superfields (and consequently of their components) in the corresponding space--time dimensions. For instance, when $\Delta=\frac 12$, in $D=3$ one finds $\frac 12$ as the canonical conformal dimension of the scalar superfield, while in the cases $D=4$  and  $D=6$, $n=8$ it is found to be equal to one and two, respectively. For convenience, we shall henceforth associate the scaling properties of the fields to the universal $D$-- and $n$--independent generalized conformal weight $\Delta$.

\subsection{Infinite--dimensional higher--spin representation of ${\mathcal N}=1$, $D=4$ supersymmetry}
Using the example of $n=4$ in $D=4$ we will now show that in four space--time dimensions, the fields of integer and half--integer spin $s=0, \frac 12, 1, \cdots, \infty$ encoded in $b(X)$ and $f_\mu(X)$
{form an irreducible infinite--dimensional supermultiplet with respect to the supersymmetry transformations generated by the \emph{ generalized} super--Poincar\'e algebra \eqref{ST}--\eqref{pX4}.} The hyperfields $b(X)$ and $f_\mu(X)$, satisfying \eqref{FF}, transform under the supertranslations \eqref{dPhi} as follows
\be\label{susybf}
\delta b(X)=-i\epsilon^\mu\,f_\mu(X)\,,\qquad \delta f_\mu(X)= -\epsilon^\nu\,\partial_{\nu\mu}\,b(X)\,.
\ee
The $D=4$ higher--spin field curvatures are contained in $b(X)$ and $f_\mu(X)$ as the components of the series expansion in the powers of the tensorial coordinates $y^{mn}$ of the flat hyperspace \eqref{X}
\begin{eqnarray}\label{ymn}
 b(x^l,\,y^{mn})=&{\hspace{-5pt}}\phi(x)+y^{m_1n_1}F_{m_1n_1}(x)
+y^{m_1n_1}\,y^{m_2n_2}\,[R_{m_1n_1,m_2n_2}(x)-{1\over 2}\eta_{m_1m_2}\partial_{n_1}\partial_{n_2}\phi(x)]\nonumber\\
&+\sum_{s=3}^{\infty}\,y^{m_1n_1}\cdots
y^{m_sn_s}\,[R_{m_1n_1,\cdots,m_sn_s}(x)+\cdots]\,,\nonumber\\
~&\\
 f^\rho(x^l,y^{mn})\equiv C^{\rho\mu}f_\mu
 =&{\hspace{-150pt}}
\psi^\rho(x)+y^{m_1n_1}[{ R}^\rho_{m_1n_1}(x)-{1\over
2}\partial_{m_1}(\gamma_{n_1}\psi)^\rho]\nonumber\\
& +\sum_{s={5\over 2}}^{\infty}\,y^{m_1n_1}\cdots y^{m_{s-{1\over
2}}n_{s-{1\over 2}}}\,[{R}^\rho_{m_1n_1,\cdots,m_{s-{1\over
2}}n_{s-{1\over 2}}}(x)+\cdots]\,.\nonumber
\end{eqnarray}
Remember that in (\ref{ymn}), $C^{\rho\mu}=-C^{\mu\rho}$ is the charge conjugation matrix used to raise spinor indices, $\phi(x)$ and $\psi^\rho(x)$ are a $D=4$ scalar and a spinor
field, respectively, $F_{m_1n_1}(x)$ is the Maxwell field strength,
$R_{m_1n_1,m_2n_2}(x)$ is the curvature tensor of linearized
gravity, ${ R}^\rho_{m_1n_1}(x)$ is the Rarita--Schwinger field
strength and other terms in the series stand for generalized Riemann
curvatures of spin--$s$ fields\footnote{The pairs of the indices separated by the commas are antisymmetrized.}
that also contain contributions of
derivatives of the fields of lower spin denoted by dots, as in the case of
the Rarita--Schwinger and gravity fields (see \cite{Bandos:2005mb} for further details).

The fact that the higher-- spin fields should form an infinite--dimensional representation {of the generalized} $ \mathcal N=1$, $D=4$ supersymmetry \eqref{ST} is prompted by the observation that the spectrum of bosonic fields contains a single real scalar field $\phi(x)$, which alone cannot have a fermionic superpartner, while each field with $s>0$  has two helicities $\pm s$. Indeed, from \eqref{susybf} we obtain an infinite entangled chain of supersymmetry transformations for the $D=4$ fields
\bea\label{susycomp}
&\delta\phi(x)=-i\epsilon^\mu\,\psi_\mu (x)\,, \qquad \delta \psi_\mu=-\frac 12 \epsilon^\nu(\gamma^m_{\nu\mu}\,\partial_m\phi+\gamma^{mn}_{\nu\mu }\,F_{mn}),& \nonumber\\
& \delta F_{mn}=-i\epsilon^\mu \left({ R}_{\mu\,mn}(x)-{1\over
2}\partial_{[m}(\gamma_{n]}\psi)_\mu\right)\,, &\\
&\delta{ R}_{\mu\,mn}(x)= {1\over
2}\partial_{[m}(\gamma_{n]}\delta \psi)_\mu-\frac 12\epsilon^\nu\,\gamma^p_{\nu\mu}\,\partial_pF_{mn} - \epsilon^\nu\,\gamma^{pq}_{\nu\mu}\left(R_{pq,mn}(x)-{1\over 2}\partial_{q}\eta_{p[m}\partial_{n]}\phi(x)\right)\,,&\nonumber
\eea
and so on.

The algebraic reason behind the appearance of the infinite--dimensional supermultiplet of the $D=4$ higher--spin fields is related to the following fact. In the $n=4$, $D=4$ case the superalgebra \eqref{ST} takes the following form
\be\label{ST4}
\{Q_\mu,Q_\nu\}=(\gamma^m)_{\mu\nu}P_m+(\gamma^{mn})_{\mu\nu} Z_{mn}\,,
\ee
where $P_m$ is the momentum along the four--dimensional space--time and $Z_{mn}=-Z_{nm}$ are the tensorial charges associated with the momenta along the extra coordinates $y^{mn}$.

On the other hand, the conventional $N=1$, $D=4$ super--Poincar\'e algebra is
\be\label{ST4p}
\{Q_\mu,Q_\nu\}=(\gamma^m)_{\mu\nu}P_m\,.
\ee
Though the both algebras have the same number of the supercharges $Q_\mu$, their anti--commutator closes on different sets of bosonic generators. So the super--Poincar\'e algebra \eqref{ST4p} is not a subalgebra of \eqref{ST4}. Hence the representations of \eqref{ST4} do not split into (finite--dimensional) representations of the standard super--Poincar\'e algebra. In this sense the supersymmetric higher--spin systems under consideration differ from the most of supersymmetric models of finite--dimensional super--Poincar\'e or AdS higher--spin supermultiplets  considered in the literature (see e.g. \cite{Curtright:1979uz,Vasiliev:1980as,Bellon:1986ki,Fradkin:1987ah,Bergshoeff:1988jm,Konstein:1989ij,
Kuzenko:1993jp,Kuzenko:1993jq,Kuzenko:1994dm,Buchbinder:1995ez,Gates:1996my,Gates:1996xs,Sezgin:1998gg,
Alkalaev:2002rq,Engquist:2002gy,Sezgin:2012ag,Zinoviev:2007js,Fotopoulos:2008ka,Gates:2013rka,Gates:2013ska, Candu:2014yva}).

It will be of interest to study which higher--spin superalgebra, associated with the enveloping algebra of $osp (1|2n)$,  underlies the super--hyperspace system under consideration. In particular, one should understand whether and how this superalgebra can be embedded into the higher--spin superalgebra $hu(1,1|2n)$ considered in \cite{Konstein:1989ij} and, in the
context of hyperspace constructions, in \cite{Vasiliev:2001zy}. For instance, in the $D=4$ case the superalgebra $hu(1,1|8)$ contains $osp(2|8)$  as a finite--dimensional subalgebra \cite{Vasiliev:2001zy}, the latter contains the $D=4$ superconformal algebra $su(1,1|4)$ and, hence, the usual $N=1$, $D=4$ super--Poincar\'e algebra as sub--superalgebras, thus allowing for an $hu(1,1|8)$--invariant higher--spin system to split into the conventional finite--dimensional $N=1$, $D=4$ supermultiplets.  As we have argued above (see also the comment in the end of Section 2.2), this is not so for the $osp(1|8)$--invariant higher--spin model under consideration. In this respect let us also note that, as has been pointed out \emph{e.g.} in \cite{Vasiliev:2004cm}, although higher--spin superalgebras exist in any space--time dimension $D$ they admit usual
finite--dimensional sub--superalgebras only in space--times of lower dimensions\footnote{The case of  $D=6$ still has to  be analyzed. We thank Mikhail Vasiliev for comments on this issue.}   such as $D=3,4,5$ and 7. In other words, higher--spin supersymmetry does not necessarily imply conventional supersymmetry.

\setcounter{equation}0
\section{Scalar superfields on $OSp(1|n)$ group manifolds and their equations of motion}\label{s3}
\subsection{Geometric structure of the $OSp(1|n)$ group manifolds}
The geometric structure of the $OSp(1|n)$ group manifolds in the form we shall review below and use extensively in this paper for the description of higher-- spin fields in the associated AdS spaces has been discussed in \cite{Bandos:1999qf,Bandos:1999pq,Plyushchay:2003gv,Plyushchay:2003tj,Bandos:2004nn}.
The $OSp(1|n)$ superalgebra is formed by $n$ anti--commuting supercharges ${\mathcal Q}_\alpha$ and $\frac{n(n+1)}{2}$ generators $M_{\alpha\beta}=M_{\beta\alpha}$ of $Sp(n)$
\bea\label{OSp}
&\{{\mathcal Q}_\alpha,{\mathcal Q}_\beta\}=2M_{\alpha\beta}\,,\qquad [{\mathcal Q}_\alpha,M_{\beta\gamma}]=\frac{i\xi}2C_{\alpha(\beta}\,{\mathcal Q}_{\gamma)},&\nonumber\\
&[M_{\alpha\beta},M_{\gamma\delta}]=-\frac{i\xi}2(C_{\gamma(\alpha}M_{\beta)\delta}+C_{\delta(\alpha}M_{\beta)\gamma})\,,&
\eea
where $C_{\alpha\beta}=-C_{\beta\alpha}$ is the $Sp(n)$ invariant symplectic metric and $\xi$ is a parameter of inverse dimension of length related to the $AdS$ radius via $r=2/\xi$ (see also \cite{Florakis:2014kfa}). The $OSp(1|n)$ algebra \eqref{OSp} is recognized as a subalgebra of \eqref{osp12n} with the identifications
\be\label{generators}
{\mathcal Q}_\alpha=(Q_\alpha+\frac\xi 4 S_\alpha), \qquad M_{\alpha\beta}=P_{\alpha\beta}-\frac {\xi^2}{16}K_{\alpha\beta}-\frac \xi 4L_{(\alpha\beta)}\,,
\ee
where $S_\alpha=S^\beta C_{\beta\alpha}$, $L_{\alpha\beta}=L_\alpha{}^\gamma C_{\gamma\beta}$ and $K_{\alpha\beta}=K^{\gamma\delta}C_{\gamma\alpha}C_{\delta\beta}$.

The $OSp(1|n)$ manifold is parametrized by the coordinates $(X^{\mu\nu},\theta^\mu)$ and its geometry is described by the Cartan forms
\be\label{Cartan}
\Omega={\mathcal O}^{-1}d{\mathcal O}(X,\theta)= -i\Omega^{\alpha\beta}M_{\alpha\beta}+iE^\alpha{\mathcal Q}_\alpha\,,
\ee
where ${\mathcal O}(X,\theta)$ is an $OSp(1|n)$ supergroup element. The Cartan forms satisfy the Maurer--Cartan equations associated with the $OSp(1|n)$ superalgebra \eqref{OSp}
\be\label{MC}
d\Omega^{\alpha\beta}+\frac \xi 2\Omega^{\alpha\gamma}\wedge \Omega_{\gamma}{}^{\beta}=-i E^\alpha \wedge E^\beta, \qquad dE^\alpha+\frac \xi 2E^{\gamma}\wedge \Omega_{\gamma}{}^{\alpha}=0\,,
\ee
with the external differential acting from the right.

The Maurer--Cartan equations \eqref{MC} are then solved by the following forms
\be\label{Omega}
\Omega^{\alpha \beta} = d X^{\mu \nu} G_\mu{}^\alpha G_\nu{}^\beta(X) + \frac{i}{2} (\Theta^\alpha {\mathcal D} \Theta^\beta +
\Theta^\beta {\mathcal D} \Theta^\alpha) = \Pi^{\mu\nu} \,{\mathcal G}_\mu{}^\alpha \, {\mathcal G}_{\nu}{}^{\beta}(X,\Theta),
\ee
\begin{eqnarray}\label{Ealpha}
E^\alpha = P(\Theta^2) {\mathcal D} \Theta^\alpha - \Theta^\alpha{\mathcal D} P(\Theta^2)\,
\end{eqnarray}
where $\Theta$ is related to $\theta$ through
\begin{equation}\label{Theta}
\theta^\alpha=\Theta^\beta G_{\beta}^{-1\alpha}P^{-1}(\Theta^2),\qquad \Theta^2=\Theta^\alpha\Theta_\alpha,\qquad P^2(\Theta^2)=1+\frac{i\xi}8\Theta^2\,,
\end{equation}
while the covariant derivative
\be\label{mathcalD}
{\mathcal D}\Theta^\alpha=d\Theta^\alpha+\frac\xi 4 \Theta^\beta\,\omega_\beta{}^\alpha(X)\,,
\ee
contains the Cartan form of the $Sp(n)$ group manifold
\be\label{omega}
\omega^{\alpha\beta}(X)=dX^{\mu\nu}G_\mu{}^\alpha(X) G_{\nu}{}^\beta(X),
\ee
and
\begin{equation}\label{calG}
{\mathcal G}_{\alpha}{}^\beta(X,\Theta)=G_{\alpha}{}^\beta(X)-\frac{i\xi}8(\Theta_\alpha-2G_{\alpha}{}^\gamma\Theta_\gamma)\Theta^\beta,\qquad G^{-1\,\beta}_\alpha=\delta^{\alpha}_\beta+\frac \xi 4 X_\alpha{}^\beta.
\end{equation}
Note also the relations
\begin{equation}\label{Theta1}
\theta^\alpha{\mathcal G}_{\alpha}{}^\beta=\Theta^\beta P(\Theta^2), \qquad \theta^\alpha=\Theta^\beta {\mathcal G}_{\beta}^{-1\alpha}P(\Theta^2)\,,
\end{equation}
and the fact that the inverse matrix of (\ref{calG}) is given by
\begin{eqnarray}\label{calGn-or1}
{\mathcal G}_{\alpha}^{-1\beta}(X,\Theta)&=&G_{\alpha}^{-1\beta}(X)
-\frac{i\xi}{8}(\Theta^\delta G_{\delta\alpha}^{-1})\,(\Theta^\delta\,G_{\delta}^{-1\beta})P^{-2}(\Theta^2)\nonumber\\
&=&G_{\alpha}^{-1\beta}(X)
-\frac{i\xi}{8}\theta_\alpha\,\theta^\beta\,=\delta_\alpha^\beta+\frac\xi 4(X_\alpha{}^\beta-\frac i2 \theta_\alpha\,\theta^\beta).
\end{eqnarray}

The form of the bosonic Cartan form \eqref{Omega} prompts us that the latter is related to the super--invariant form \eqref{VA} in flat hyper superspace via the $GL(n)$ transformation with matrix element \eqref{calG}. This property was revealed in \cite{Plyushchay:2003gv} and called GL--flatness of the $OSp(1|n)$ supermanifold. It will allow us to generalize the results of \cite{Florakis:2014kfa} and relate the scalar superfield $\Phi(X,\theta)$ and its field equation \eqref{DDPhi} in flat superspace to a scalar superfield and its equation of motion on the supergroup manifold $OSp(1|n)$.
\subsection{Scalar superfield on $OSp(1|n)$ and its $OSp(1|2n)$ invariant equation of motion}

The scalar superfield equation on $OSp(1|n)$ takes the form \cite{Bandos:2004nn}
\be\label{Osp}
\left({\nabla}_{[\alpha }{\nabla}_{\beta ]}-\frac{i\xi}8C_{\alpha \beta }\right)\Phi_{OSp}(X,\theta)=0\,,
\ee
where the Grassmann--odd covariant derivatives $\nabla_\alpha$ and their bosonic counterparts $\nabla_{\alpha\beta}$ satisfy the $OSp(1|n)$ superalgebra similar to \eqref{OSp}, namely
\be\label{antic}
\{{\nabla}_{\alpha},{\nabla}_{\beta}\}=2i \nabla_{\alpha\beta}\,
\ee
\be\label{vs}
[\nabla_\gamma,\nabla_{\alpha\beta}]=\frac{\xi}2C_{\gamma(\alpha}\,\nabla_{\beta)},
\ee
\begin{equation} \label{algebra}
[\nabla_{\alpha \beta}, \nabla_{\gamma \delta}] = \frac{\xi}{2}(C_{\alpha (\gamma} \nabla_{\delta)\beta}+ C_{\beta (\gamma} \nabla_{\delta)\alpha }
)\,.
\end{equation}
A somewhat tedious but straightforward algebra then shows that the superfield $\Phi_{OSp}(X,\theta)$ satisfying \eqref{Osp} is related to the superfield $\Phi(X,\theta)$ satisfying the flat superspace equation \eqref{DDPhi} by the super--Weyl transformation
\begin{equation}\label{W}
\Phi_{OSp(1|n)}(X,\Theta)=({\det {\mathcal G}})^{-\frac 12}\,\Phi_{flat}(X,\theta)=({\det {G}})^{-\frac 12}P(\Theta^2)\,\Phi_{flat}(X,\theta),
\end{equation}
while the $OSp(1|n)$ covariant derivatives are obtained from the flat superspace ones by the following GL (`generalized superconformal') transformations
\bea\label{nablaD}
\nabla_\alpha&=&{\mathcal G}_{\alpha}^{-1\,\mu}\,D_\mu\,,\nonumber\\
\nabla_{\alpha\beta}&=&{\mathcal G}_{\alpha}^{-1\,\mu}\,{\mathcal G}_{\beta}^{-1\nu}\left(\partial_{\mu\nu}+2i{D}_{(\mu}\ln\left((\det G)^{\frac 12}P^{-1}(\Theta^2)\right)\,{D}_{\nu)}\right).
\eea
Substituting \eqref{bftheta} into \eqref{W} and using the definition (\ref{Theta}), together with the fact that on the mass shell all higher components in \eqref{bftheta} vanish, we find
\begin{equation}\label{composp}
\Phi_{OSp(n)}(X,\Theta)=({\det {G}})^{-\frac 12}\,b(X)+\Theta^\alpha ({\det {G}})^{-\frac 12}\,G_\alpha^{-1\mu}(X) \,f_\mu(X) +O(\Theta^2,b(X)),
\end{equation}
where the first two terms are the fields
\be\label{BF1}
B(X)=({\det {G}})^{-\frac 12}\,b(X),\qquad F_\alpha(X)=({\det {G}})^{-\frac 12}\,G_\alpha^{-1\mu}(X) \,f_\mu(X)
\ee
propagating on the $Sp(n)$ group manifold, and $O(\Theta^2,b(x))$ stands for higher order terms in $\Theta^2$ which only depend on $b(X)$. The fields \eqref{BF1} satisfy the equations of motion
\begin{eqnarray} \label{BADS}
&&(\nabla_{\alpha \beta} \nabla_{\gamma \delta}- \nabla_{\alpha \gamma} \nabla_{\beta \delta})B- \\ \nonumber
&&-\frac{\xi}{8}(C_{\alpha \gamma} \nabla_{\beta \delta}  - C_{\alpha \beta } \nabla_{\gamma \delta} + C_{\beta \delta} \nabla_{\alpha \gamma}- C_{\gamma \delta}\nabla_{\alpha \beta} + 2 C_{\beta \gamma} \nabla_{\alpha \delta} ) B-\\ \nonumber
&&  -(\tfrac{\xi}{8})^2 (C_{\alpha\gamma}C_{\beta\delta} - C_{\alpha \beta}C_{\gamma \delta}  + 2C_{\beta\gamma}C_{\alpha\delta})
B=0\,,
\end{eqnarray}
\begin{equation}\label{FADS}
\nabla_{\alpha \beta} F_\gamma - \nabla_{\alpha \gamma} F_\beta +\frac{\xi}{8}
   (C_{ \gamma \alpha} F_\beta - C_{ \beta \alpha} F_\gamma + 2 C_{\gamma \beta} F_\alpha    )=0\,,
 \end{equation}
discussed in detail in \cite{Florakis:2014kfa}. Note that in \eqref{BADS} and \eqref{FADS} the covariant derivatives are restricted to the bosonic group manifold $Sp(n)$, \emph{i.e.} $\nabla_{\alpha\beta}=G^{-1\,\mu}_\alpha(X)\,G^{-1\,\nu}_\beta(X)\,\partial_{\mu\nu}$.

Since the flat superspace field equation is invariant under the generalized superconformal $OSp(1|2n)$ transformations \eqref{deltaPhi}, the above relation leads us to conclude that also the $OSp(1|n)$ superspace equations \eqref{Osp} are invariant under the $OSp(1|2n)$ transformations, under which the superfield $\Phi_{OSp}(X,\theta)$ varies as
\bea\label{deltaPhi-osp}
\delta\Phi_{OSp} &=&-(\epsilon^\mu\,{\mathbb Q}_\mu+\xi_\mu\, {\cal S}^\mu+ia^{\mu\nu}\,{\cal P}_{\mu\nu}+ik_{\mu\nu}\,{\cal K}^{\mu\nu}+ig_\mu{}^\nu\,{\cal L}_{\nu}{}^\mu)\,\Phi_{OSp}\nonumber\\
&&-\frac 12\, \left(g_\mu{}^\mu-
 k_{\mu \nu} (X^{\mu \nu}+\frac i2 \theta^\mu\theta^\nu)+\xi_\mu\,\theta^\mu\right)\,\Phi_{OSp}\,.
\eea
Here
\be\label{KP}
{\cal P}_{\mu \nu}= - i{\cal D_{\mu \nu}} = -i (\partial_{\mu \nu} + \frac{\xi}{8} \cal G_{(\alpha \beta)} )\,,
\ee
and
\be\label{KQ}
{\mathbb Q}_\mu= Q_\mu - \frac{i \xi}{8} \Theta_\mu P(\Theta)\,.
\ee
Using the relations
\be\label{2-1}
Q_\beta\Theta^\alpha=P^{-1}(\Theta^2)\left(G_\beta{}^\alpha +\frac{i\xi}8
\Theta_\beta \Theta^\alpha +
\frac{i\xi}8 G_\beta{}^\sigma
\Theta_\sigma \Theta^\alpha
+
\left(\frac{i\xi}8\right)^2 \Theta^2 \Theta_\beta\Theta^\alpha\right)\,,
\ee
\be\label{3-1}
(Q_\beta\Theta^\alpha)\Theta_\alpha=P(\Theta^2)\left(G_\beta{}^\sigma
 +\frac{i\xi}{8}\Theta_\beta\Theta^\sigma\right)\Theta_\sigma,
\ee
\be \label{4-1}
\partial_{\alpha\beta}\Theta^\gamma =\frac\xi 4\Theta_{(\alpha}G_{\beta)}{}^\delta(\delta_\delta^\gamma+\frac{i\xi}8\Theta_\delta\Theta^\gamma)\,,
\ee
\be\label{5-1}
D_\beta{\mathcal G}_\alpha{}^\gamma =\frac{i\xi}4P(\Theta^2)\,(\Theta_\alpha-2G_\alpha{}^\rho\Theta_\rho){\mathcal G}_\beta{}^\gamma
\ee
\be\label{6-1}
\partial_{\alpha\beta}{\mathcal G}_\gamma{}^\delta=\frac \xi 4{\mathcal G}_{\gamma(\alpha}\,{\mathcal G}_{\beta)}{}^\delta\,,
\ee
and
\be \label{7-1}
Q_\alpha {\mathcal G}_{\mu \nu} =  -\frac{i\xi}{4} P(\Theta^2) \Theta_\nu {\mathcal G}_{\mu \alpha}\,,
\ee
one may check that
the operators (\ref{KP}) and (\ref{KQ}) obey the flat hyperspace supersymmetry algebra
\be \label{KALG}
[{\cal P}_{\mu \nu}, {\cal P}_{\rho \sigma}] =0, \qquad \{ {\mathbb Q}_{\mu }, {\mathbb Q}_{\nu}  \} =- 2 {\cal P}_{\mu \nu},
\qquad [{\cal P}_{\mu \nu}, {\mathbb Q}_{\rho }] =0 \,.
\ee

The other generators of the $OSp(1|2n)$ are
\be\label{KS}
{\cal S}^\mu=-(X^{\mu\nu}+\frac i2\theta^\mu\theta^\nu){\mathbb Q}_\nu\,, \quad
{\cal L}_{\mu}{}^{\nu}=-{2i}(X^{\nu\rho}+\frac i2 \theta^\nu\theta^\rho){\cal D}_{\rho\mu}-i\theta^\nu\,{\mathbb Q}_\mu\,,
\ee
and
\be\label{KK}
{\cal K}^{\mu\nu}=i(X^{\mu\rho}+\frac i2\theta^\mu\theta^\rho)(X^{\nu\lambda}+\frac i2\theta^\nu\theta^\lambda){\cal D}_{\rho\lambda}-i\theta^{(\mu}{\cal S}^{\nu)}\,.
\ee
Taking into account  the commutation relations (\ref{KALG}) we see that
the operators ${\mathbb Q}_\mu, {\cal S}^\mu, {\cal P}_{\mu \nu}, {\cal L}_{\mu}{}^{\nu}, {\cal K}^{\mu\nu}  $
obey the same $OSp(1|2n)$ algebra \eqref{osp12n} as the operators
${ Q}_\mu, { S}^\mu, { P}_{\mu \nu}, { L}_{\mu}{}^{\nu}$ and $ K^{\mu\nu} $.

\section{Correlation functions in ${\mathcal N}=1$, $D=3$ superconformal models} \label{s4} \setcounter{equation}0

Before considering  correlation functions for  superfields in hyper superspaces,
it is instructive to discuss in detail analogous structures arising in the
superconformal theory of a real scalar superfield in a conventional ${\mathcal N}=1$, $D=3$ superspace.
The reason being that this model is the simplest example (with $n=1$)
of the $OSp(1|2n)$ invariant systems considered above.
The physical content of this system is a real scalar and a $D=3$ Majorana spinor field whereas the massless higher-- spin fields are absent.

The superconformally invariant two-- and three--point correlation functions of the ${\mathcal N}=1$, $D=3$
model have been constructed in \cite{Park:1999cw} with the use of a slightly different notation.
Below we shall discuss properties of the
two-- and three--point functions for the $D=3$ scalar superfield and its components using a formalism which straightforwardly generalizes to higher--dimensional hyperspaces.

Let us use the spinor--tensor representation for the description of the three--dimensional space--time coordinates
\be\label{x3}
x^{\alpha \beta}= x^{\beta \alpha} = x^m (\gamma_m)^{\alpha \beta},
\ee
where $\alpha, \beta=1,2$ are $D=3$ spinorial indices and $m=0,1,2$ is the vectorial one. Since \eqref{x3} provides a representation of the symmetric $2 \times 2 $ matrices $x^{\alpha\beta}$, no extra coordinates, like $y^{mn}$, are present and, hence, no higher-- spin fields.

The inverse matrix  of \eqref{x3}, $x^{-1}_{\alpha\beta}$
\be
x^{\alpha \beta} \,x^{-1}_{\beta \gamma} = \delta_\alpha^\gamma\,,
\ee
takes the simple form
\be
x^{-1}_{\alpha \beta} = -\frac{1}{x^m x_m} x^n (\gamma_n)_{\alpha \beta} =- \frac{1}{x^2} x_{\alpha \beta}\,.
\ee
We may now consider a real scalar  superfield in $D=3$
\be \label{superfield}
\Phi(x, \theta) = \phi(x) + i \theta^\alpha f_\alpha (x) + \theta^\alpha \theta_\alpha F(x)\,,
\ee
with $ \phi(x)$ being a physical scalar, $f_\alpha(x)$  a physical fermion and $F(x)$ an auxiliary field.

If \eqref{superfield} satisfies the free equation of motion \eqref{DDPhi}, which in the $D=3$ case reduces to
\be\label{DDPhi3}
D^\alpha D_\alpha \Phi=0\,,
\ee
 the auxiliary field $F(x)$ vanishes, the scalar field $\phi(x)$ satisfies the massless Klein--Gordon equation and
$f_\alpha(x)$ satisfies the massless Dirac equation.

Let us consider a superconformal transformation of\eqref{superfield}.
The Poincar\'e supersymmetry transformations read
\be \label{PS}
\delta \Phi(x, \theta) = \epsilon^\alpha \left( \frac{\partial}{ \partial \theta^\alpha} - i  \theta^{ \beta}\frac{\partial}{ \partial x^{\alpha \beta}} \right) \Phi(x, \theta) =  \epsilon^\alpha  Q_\alpha \Phi(x, \theta)\,,
\ee
and imply the supersymmetry transformations of the component fields
\begin{eqnarray}
&&\delta \phi (x) = i \epsilon^\alpha f_\alpha(x)\,, \\
&&\delta f_\alpha(x) = -2i \epsilon_\alpha F(x)  - \epsilon^\beta \partial_{\alpha \beta} \phi(x)\,, \\
&& \delta F(x) = \frac{1}{2} \epsilon^\alpha \partial_{\alpha \beta} f^\beta(x)\,,
\end{eqnarray}
where we have made use of the identity
\be
\theta^\alpha \theta^\beta =\frac{1}{2} C^{\alpha \beta} (\theta^\gamma \theta_\gamma)\,.
\ee
Moreover, under conformal supersymmetry, $\Phi(x, \theta)$ transforms as
\be \label{CS}
\delta \Phi(x, \theta)  = \xi_\alpha (x^{\alpha \beta} + \frac{i}{2} \theta^\alpha \theta^\beta) Q_\beta  \Phi(x, \theta)
- i (\xi_\alpha \theta^\alpha) \Delta \Phi(x, \theta) \,,
\ee
where $\Delta$ is the conformal weight of the superfield. The superconformal transformations of the component fields are given by
\begin{eqnarray}
&&\delta \phi(x)= i \xi_\alpha\,\, x^{\alpha \beta} f_\beta(x), \\
&& \delta f_\alpha(x)= -2i \xi_\beta \,\, x^\beta{}_\alpha F(x) + \xi_\beta \,\, x^{\beta \gamma}\,\, \partial_{\gamma \alpha} \phi(x)
+ \xi_\alpha \Delta \phi(x), \\
&& \delta F(x) = \frac{1}{2}\,\, \xi_\alpha \,\, x^{\alpha \beta} \partial_{\beta \gamma} f^\gamma(x) - \frac{1}{2} \xi_\alpha
\left( \frac{1}{2} - \Delta \right)  f^\alpha(x).
\end{eqnarray}
The conformal weights of $\phi$, $f_\alpha$ and $F$ are $\Delta$, $\Delta +\frac 12$ and $\Delta +1$, respectively.

It should be noted that the field equation \eqref{DDPhi3} is superconformally invariant if the superfield $\Phi(x,\theta)$ has the canonical conformal weight $\Delta=\frac 12$.

\subsection{Two--point functions} \label{Tpf-3d-n1}
The form of correlation functions in superconformal theories is drastically restricted by the requirement of their superconformal invariance.

The two--point correlation function of the superfield $\Phi(x,\theta)$ with conformal weight $\Delta$ is obtained by first solving the superconformal Ward identities which involve $Q$-- and $S$--supersymmetry transformations. The invariance under bosonic translations, rotations, conformal boosts and dilations then follows as a consequence of the properties of the superconformal algebra. The $Q$-- and $S$--supersymmetry Ward identities are
\be \label{WI2PT-1}
\epsilon^\mu \left( \frac{\partial}{\partial \theta_1^\mu}  - i \theta_1^\nu \frac{\partial }{\partial x_1^{\mu \nu}}
+
\frac{\partial}{\partial \theta_2^\mu}  - i \theta_2^\nu\frac{\partial }{\partial x_2^{\mu \nu}}
\right)\langle \Phi(x_1, \theta_1) \Phi(x_2, \theta_2) \rangle =0\,,
\ee
and
\bea \nonumber
&&\xi_\mu \left[ ( X_1^{\mu \nu} + \frac{i}{2} \theta_1^\mu \theta_1^\nu )
 \left ( \frac{\partial}{\partial \theta_1^\nu}  - i \theta_1^\rho \frac{\partial }{\partial x_1^{\nu \rho}} \right ) +
( X_2^{\mu \nu} + \frac{i}{2} \theta_2^\mu \theta_2^\nu )
 \left ( \frac{\partial}{\partial \theta_2^\nu}  - i \theta_2^\rho \frac{\partial }{\partial x_2^{\nu \rho }} \right ) \right ]  \\ \nonumber
&& \cdot \langle \Phi(x_1, \theta_1) \Phi(x_2, \theta_2) \rangle
 +i \Delta \,\xi_\mu  \left (   \theta^\mu_1 +  \theta^\mu_2 \right)
 \langle \Phi(x_1, \theta_1) \Phi(x_2, \theta_2) \rangle =0 \,.
\eea
The solution to these equations takes the form
\be \label{2pt-g}
\langle \Phi(x_1, \theta_1) \Phi(x_2, \theta_2)   \rangle=c_2
 ({\rm det} |z_{12}|)^{-\Delta} \,,
\ee
where $c_2$ is an arbitrary normalization constant and
\be\label{sinter}
z_{ij}^{\mu \nu} = x_i^{\mu \nu} - x_j^{\mu \nu} - \frac{i}{2} \theta_i^\mu \theta_j^\nu -
\frac{i}{2} \theta_i^\nu \theta_j^\mu\,,
\ee
is invariant under $Q$--supersymmetry.  As usual, for the two--point function to be non--vanishing, the
conformal weights of the two superfields should be equal.

Expanding  the expression on the right hand side of (\ref{2pt-g}) in powers of $\theta$, we obtain
\bea \label{expansion-2}
	\begin{split}
({\rm det} |z_{12}|)^{-\Delta}&= ({\rm det} |x_{12}|)^{-\Delta} -i \partial_{\alpha \beta}  ({\rm det} |x_{12}|)^{-\Delta}\,
\theta_1^{( \alpha} \theta_2^{\beta )} \\
	&- \frac{1}{2}
\partial_{\gamma \delta}
\partial_{\alpha \beta}  ({\rm det} |x_{12}|)^{-\Delta}\,\theta_1^{( \alpha} \theta_2^{\beta )}\theta_1^{( \gamma} \theta_2^{\delta )} \,.
	\end{split}
\eea
Using the identities
\be
\partial_{\alpha \beta}( {\rm det} |x|)^{-\Delta} = -\Delta\, x^{-1}_{\alpha \beta}\,\,\, {\rm det}| x|^{-\Delta} \,,
\ee
and
\be
\partial_{\alpha \beta} \partial_{\gamma \delta}  ({\rm det} |x|)^{-\Delta} =
\Delta \left( \Delta \,x^{-1 }_{\alpha \beta} x^{-1 }_{\gamma \delta}    + \frac{1}{2} x^{-1 }_{\alpha \gamma} x^{-1 }_{\beta \delta} +
\frac{1}{2} x^{-1 }_{\beta \gamma} x^{-1 }_{\alpha \delta} \right)
 ({\rm det} |x|)^{-\Delta} \,,
\ee
 one may rewrite  the expression  (\ref{expansion-2}) as
\be \label{expansion-3-1}
(\det |z_{12}|)^{-{\Delta}}=
(\det |x_{12}|)^{-\Delta} \left(1 - {i \Delta} \frac{x^m_{12} (\gamma_m)_{\alpha \beta}}{x_{12}^2}
\theta_1^{\alpha} \theta_2^{\beta } - \frac{(2\Delta -1) \Delta}{4}
\frac{1}{ x_{12}^2} \theta_1^2 \theta_2^2 \right).
\ee
Thus, from  equations (\ref{expansion-2}) or \eqref{expansion-3-1}, one may immediately read off the expressions for the correlation functions of the component fields of the superfield
(\ref{superfield}).
\be
\langle \phi(x_1) \phi(x_2) \rangle =c_2({\rm det} |x_{12}|)^{-\frac{1}{2}}\,,
\ee

\be
\langle f_\alpha(x_1) f_\beta(x_2) \rangle =- ic_2 \partial_{\alpha \beta}({\rm det} |x_{12}|)^{-\frac{1}{2}}\,, \qquad \langle \phi(x_1) f_\alpha(x_2) \rangle =0\,,
\ee

\be\label{F}
 \langle F(x_1) \phi(x_2) \rangle =0\,, \qquad \langle F(x_1) f_\alpha(x_2) \rangle =0\,.
\ee

\be \label{FFa}
\langle F(x_1) F(x_2) \rangle = -\frac{c_2}{8}  \partial^{\alpha \beta } \partial_{\alpha \beta}  ({\rm det} |x|)^{-\Delta} \,.
\ee

Let us note that when the superfield $\Phi(x,\theta)$ has the canonical conformal dimension $\Delta=\frac{1}{2}$,
 due to the identity
\be
C^{\alpha \gamma} C^{\beta \delta} \partial^1_{\alpha \beta} \partial^1_{\gamma \delta }
({\rm det} |x_{12}|)^{-\frac{1}{2}}
= - \frac{1}{2} \eta^{mn}\frac\partial{\partial x_1^m}\frac\partial{\partial x_1^n} ({\rm det} |x_{12}|)^{-\frac{1}{2}}
\ee
the last term  in (\ref{expansion-2})
is proportional to the $\delta$--function if one moves to the Euclidean signature.
Then one has for the two--point function for the auxiliary field
\be \label{FF-1}
\langle F(x_1) F(x_2) \rangle = -\frac{\pi }{4} c_2 \delta^{(3)} (x_1-x_2).
\ee
Note that the correlation functions of the auxiliary field $F$ with the
 physical fields and with itself (for $x^m_1 \neq x^m_2$) vanish.

On the other hand, if the conformal weight of the superfield (\ref{superfield}) is anomalous, i.e.
$\Delta \neq \frac{1}{2}$, the correlators of the auxiliary field with the physical ones still vanish (in agreement with the fact that their conformal weights are different), but the $\langle F F \rangle $ correlator is
\be\label{Fa}
\langle F(x_1) F(x_2) \rangle =- c_2 \frac{(2\Delta -1) \Delta}{4}
\frac{1}{ x_{12}^2}\,(\det |x_{12}|)^{-\Delta}=- c_2 \frac{(2\Delta -1) \Delta}{4}(\det |x_{12}|)^{-\Delta-1}.
\ee
This situation may correspond to an interacting quantum ${\mathcal N}=1$ superconformal field theory \cite{Synatschke:2010ub}, where the auxiliary field is non--zero, and fields acquire anomalous dimensions due to quantum corrections.

\subsection{Three--point functions}  \label{3pf-3d-n1}

We now  consider three--point functions involving three real scalar superfields carrying scaling dimensions $\Delta_i$ ($i$=1,2,3).
Solving the superconformal Ward identities for $Q$-- and $S$--supersymmetry transformations we find
\be \label{3p-1}
 \langle \Phi^{\Delta_1}(x_1, \theta_1)  \Phi^{\Delta_2}(x_2, \theta_2)   \Phi^{\Delta_3}(x_3, \theta_3) \rangle= c_3(\det |z_{12}|)^{-k_1} (\det |z_{23}|)^{-k_2} (\det |z_{31}|)^{-k_3}\,,
\ee
where
\be\label{ki}
k_1 = \frac{1}{2} (\Delta_1 + \Delta_2 - \Delta_3)\,, \quad
k_2 = \frac{1}{2} (\Delta_2 + \Delta_3 - \Delta_1)\,, \quad
k_3 = \frac{1}{2} (\Delta_3 + \Delta_1 - \Delta_2)\,.
\ee
Using the expansion (\ref{expansion-3-1}),  one obtains the three--point functions of the component
fields of $\Phi^{\Delta_1}(x_1, \theta_1) $, $\Phi^{\Delta_2}(x_2, \theta_2)$ and $\Phi^{\Delta_3}(x_3, \theta_3)$, whose labels of  scaling dimension we skip for simplicity
\be \label{3p-21}
 \langle \phi(x_1)  \phi(x_2)   \phi(x_3) \rangle= c_3(\det |x_{12}|)^{-k_1} (\det |x_{23}|)^{-k_2} (\det |x_{31}|)^{-k_3}\,,
\ee
\bea \label{3p-3-1}
&& \langle f_\alpha(x_1) f_\beta (x_2)   \phi (x_3) \rangle = \\ \nonumber
&&=- i c_3 \frac{ k_1 x^m_{12} (\gamma_m)_{\alpha \beta}}{ x^2_{12}}
  (\det |x_{12}|)^{-k_1} (\det| x_{23}|)^{-k_2} (\det |x_{31}|)^{-k_3}\nonumber\\
 &&=- i c_3{ k_1 x^m_{12} (\gamma_m)_{\alpha \beta}}
  (\det |x_{12}|)^{-k_1-1} (\det| x_{23}|)^{-k_2} (\det |x_{31}|)^{-k_3}\,,
\eea
\bea \nonumber
&& \langle f_\alpha(x_1)    F(x_2)   f_\beta(x_3) \rangle= \\
&&=c_3
\frac{k_1 k_2}{2 x^2_{12} x^2_{23}} (\gamma_m)_\alpha{}^\delta (\gamma_n)_{\delta \beta} (x_{12}^m) (x_{23}^n)
(\det |x_{12}|)^{-k_1} (\det |x_{23}|)^{-k_2} (\det |x_{31}|)^{-k_3}\nonumber\\
&&=c_3
\frac{k_1 k_2}{2 } (\gamma_m)_\alpha{}^\delta (\gamma_n)_{\delta \beta} (x_{12}^m) (x_{23}^n)
(\det |x_{12}|)^{-k_1-1} (\det |x_{23}|)^{-k_2-1} (\det |x_{31}|)^{-k_3}\,.
\eea
\be\label{3p-4-1}
 \langle F(x_1)  F(x_2)   \phi(x_3) \rangle =- \frac{c_3}{8}
\partial^m \partial_m((\det |x_{12}|)^{-k_1}) (\det| x_{23}|)^{-k_2} (\det |x_{31}|)^{-k_3}
\ee

The remaining three--point functions  containing an odd number of fermions, as well as the correlator $\langle F\phi\phi\rangle$, vanish.
Note that, dimensional arguments would allow for a non--zero $\langle F\phi\phi\rangle$ correlator, but supersymmetry forces it to vanish. The correlator  $ \langle F(x_1)  F(x_2)   F(x_3) \rangle$ is zero as well, since it is proportional to
$(\gamma_m \gamma_n \gamma_p)x_{12}^m x_{23}^n x_{31}^p =2i \epsilon_{mnp}x_{12}^m x_{23}^n x_{31}^p=0.$

Moreover, from the above expressions we see that superconformal symmetry does not fix the values of the scaling dimensions $\Delta_i$ \eqref{ki} entering the right hand side of \eqref{3p-1}. This indicates that quantum operators may acquire anomalous dimensions and the quantum ${\mathcal N}=1$, $D=3$ superconformal theory of scalar superfields can be non--trivial, in agreement \emph{e.g.} with the results of \cite{Synatschke:2010ub}.

If the value of $\Delta$ were restricted by superconformal symmetry to its canonical value and no anomalous dimensions were allowed (for all the operators which are not protected by supersymmetry) one
would conclude that the conformal fixed point is that of the free theory. This is the case, for instance, for the ${\mathcal N}=1$, $D=4$ Wess--Zumino model in which the chirality of ${\mathcal N}=1$ matter multiplets and their three--point functions restricts the scaling dimensions of the chiral scalar supermultiplets to be canonical. This implies that in the conformal fixed point the coupling constant is zero, \emph{i.e.} the theory is free \cite{Ferrara:1974fv,Conlong:1993eu}.

\section{Correlation functions in $OSp(1|2n)$--invariant models} \label{s5} \setcounter{equation}0
Following the example of the  ${\mathcal N}=1$, $D=3$  superconformally invariant model
of the previous section,
we now proceed to compute correlation functions on hyper superspace for generic $OSp(1|2n)$ invariant models. Again, it is sufficient
to  require the invariance of the correlation functions under  $Q$-- and $S$--supersymmetry transformations. The invariance under the generalized translations, rotations and conformal transformations will then be guaranteed by the form of the $OSp(1|2n)$ superalgebra. As we will see, the form of the super--correlators will be exactly the same as in the $D=3$ case with only difference that the superinvariant intervals \eqref{sinter} are now $n \times n$ matrices.

\subsection{Two--point functions}

Let us denote the two-point correlation function by
\be
W(Z_1,Z_2)= \langle \Phi(X_1, \theta_1)  \Phi(X_2, \theta_2)   \rangle \,.
\ee
The invariance under $Q$--supersymmetry requires
\be
\epsilon^\mu \left( \frac{\partial}{\partial \theta_1^\mu}  - i \theta_1^\nu \frac{\partial }{\partial X_1^{\mu \nu}}
+
\frac{\partial}{\partial \theta_2^\mu}  - i \theta_2^\nu\frac{\partial }{\partial X_2^{\mu \nu}}
\right)W(Z_1, Z_2)=0 \,,
\ee
which implies
\be
 \langle \Phi(X_1, \theta_1)  \Phi(X_2, \theta_2)   \rangle= W({\rm det} |Z_{12}|),
\ee
where
\be
Z_{12}^{\mu \nu} = X_1^{\mu \nu} - X_2^{\mu \nu} - \frac{i}{2} \theta_1^\mu \theta_2^\nu -
\frac{i}{2} \theta_1^\nu \theta_2^\mu \,
\ee
is the interval between two points in hyper--superspace which is invariant under the rigid supersymmetry transformations \eqref{susyv}.

We next impose invariance of the correlator under the $S$--supersymmetry transformation
\bea \nonumber
&&\xi_\mu \left[  ( X_1^{\mu \nu} + \frac{i}{2} \theta_1^\mu \theta_1^\nu )
 \left ( \frac{\partial}{\partial \theta_1^\nu}  - i \theta_1^\rho \frac{\partial }{\partial X_1^{\nu \rho}} \right ) +
( X_2^{\mu \nu} + \frac{i}{2} \theta_2^\mu \theta_2^\nu )
 \left ( \frac{\partial}{\partial \theta_2^\nu}  - i \theta_2^\rho \frac{\partial }{\partial X_2^{\nu \rho }} \right ) \right ] W({\rm det} |Z_{12}|)  \\
&& + \xi_\mu \left (  \frac{i}{2} \theta^\mu_1 + \frac{i}{2} \theta^\mu_2 \right)  W({\rm det} |Z_{12}|)=0 \,,
\eea
which is solved by
\be \label{2PTF}
 W({\rm det} |Z_{12}|)= c_2 ({\rm det} |Z_{12}|)^{-\frac{1}{2}} \quad \Rightarrow \quad \langle \Phi(X_1, \theta_1)  \Phi(X_2, \theta_2)   \rangle= c_2
 ({\rm det} |Z_{12}|)^{-\frac{1}{2}}\,.
\ee
The two--point function (\ref{2PTF})   reproduces
the correlators of the component bosonic and fermionic hyperfields $b(X)$ and $f_\mu(X)$
after the expansion of the former in powers of the Grassmann coordinates $\theta_1^{( \mu} \theta_2^{\nu )}$.
Since on the mass shell the superfield (\ref{bftheta}) has only two non--zero components,
all terms  in the $\theta$-expansion  of the two-point function  (\ref{2PTF}),
starting from the ones quadratic  in  $\theta_1^{( \mu} \theta_2^{\nu )}$, should vanish.
This is indeed the case, as a consequence of the field equations.

To see this,  let us recall that in the separated points the two--point function of the bosonic hyperfield of weight $\frac 12$ satisfies the free field
equation. Therefore for $X^1_{\alpha \beta} \neq X^2_{\alpha \beta}$ one has\footnote{When the two points coincide, one can define an analog
of the Dirac delta function in the tensorial spaces, see \cite{Vasiliev:2001dc} for the relevant discussion.}
 \be \label{2ptch}
(\partial^1_{\mu \nu} \partial^1_{\rho \sigma }  - \partial^1_{\mu \rho} \partial^1_{\nu \sigma } )
\langle b(X_1) b(X_2) \rangle=
(\partial^1_{\mu \nu} \partial^1_{\rho \sigma }  - \partial^1_{\mu \rho} \partial^1_{\nu \sigma } )
({\rm det} |X_{12}|)^{-\frac{1}{2}}=0 \,.
\ee
Similarly, for $X^1_{\alpha \beta} \neq X^2_{\alpha \beta}$ the fermionic two--point function satisfies the free field equation for the fermionic hyperfield.
Written in terms of  the superfields,  these equations are encoded in the superfield equation
\be\label{EOM2}
(D^1_{\mu } D^1_{\nu }  - D^1_{\nu} D^1_{\mu } )
\langle \Phi(X_1, \theta_1)  \Phi(X_2, \theta_2) \rangle=
(D^1_{\mu } D^1_{\nu }  - D^1_{\nu} D^1_{\mu } )
({\rm det} |Z_{12}|)^{-\frac{1}{2}}=0\, \quad ({\rm for}~~Z_{12}\not =0).
\ee
Expanding the two--point function $({\rm det} |Z_{12}|)^{-\frac{1}{2}}$ in powers of
the Grassmann theta--variables
\bea \label{expansion}
	\begin{split}
&({\rm det} |Z_{12}|)^{-\frac{1}{2}}= \\
&({\rm det} |X_{12}|)^{-\frac{1}{2}} -i \partial_{\alpha \beta}  ({\rm det} |X_{12}|)^{-\frac{1}{2}}
\theta_1^{( \alpha} \theta_2^{\beta )} - \frac{1}{2}
\partial_{\gamma \delta}
\partial_{\alpha \beta}  ({\rm det} |X_{12}|)^{-\frac{1}{2}}\theta_1^{( \alpha} \theta_2^{\beta )}\theta_1^{( \gamma} \theta_2^{\delta )}
+ \ldots\,,
	\end{split}
\eea
 one may see that terms in the expansion starting from
$(\theta_1^{( \mu} \theta_2^{\nu )})^2$ vanish due to the free field equation
(\ref{2ptch}).
From equations   (\ref{2PTF}),   (\ref{expansion}) and from the explicit form of the superfield
(\ref{bftheta}), one may immediately reproduce the correlation functions for the component fields
\cite{Vasiliev:2003jc}
\be
\langle b(X_1) b(X_2) \rangle = c_2 ({\rm det} |X_{12}|)^{-\frac{1}{2}}\,, \quad
\langle f_\mu(X_1) f_\nu (X_2) \rangle = \frac{ic_2}{2} (X_{12})^{-1}_{\mu \nu} ({\rm det} |X_{12}|)^{-\frac{1}{2}}\,.
\ee
Notice also that, contrary to the non--supersymmetric case, where the two--point functions for bosonic and fermionic hyperfields
contain an independent normalization constant each, in the supersymmetric case the number of independent constants
is reduced to one.

The two--point functions on the $OSp(1|n)$ manifold may now be obtained from
(\ref{2PTF}) via the rescaling \eqref{W}, which relates the superfields in flat superspace and on the $OSp(1|n)$ group manifold
\bea\label{osp2p}
&&\langle \Phi_{OSp}(X_1, \theta_1)  \Phi_{OSp}(X_2, \theta_2)   \rangle = \\ \nonumber
&& ({\rm det}\, G (X_1))^{-\frac{1}{2}} P(\Theta_1^2) ({\rm det}\, G (X_2))^{-\frac{1}{2}} P(\Theta_2^2)
\langle \Phi(X_1, \theta_1)  \Phi(X_2, \theta_2)\rangle\,.
\eea
Finally, as in the $D=3$ case, one may derive the superconformally invariant two--point function for superfields carrying an arbitrary generalized conformal weight $\Delta$, which on flat hyper superspace has the form
\be\label{2pd}
\langle \Phi^{\Delta_1}(X_1, \theta_1)  \Phi^{\Delta_2}(X_2, \theta_2)   \rangle= c_2
 ({\rm det} |Z_{12}|)^{-\Delta}\,, \qquad \Delta_1=\Delta_2=\Delta\,.
 \ee
In principle, in order to obtain the $OSp(1|n)$ correlator, as in the case $\Delta=\frac 12$, one may apply to \eqref{2pd} a Weyl rescaling similar to \eqref{osp2p}. However, when $\Delta \not =\frac 12$  the superfields no longer satisfy the quadratic equations \eqref{DDPhi} and \eqref{Osp}, because the latter equations are superconformally invariant only for $\Delta =\frac 12$. Thus, fixing  the power of $({\rm det} G (X))^{-\frac{1}{2}} P(\Theta^2)$ in the Weyl transform of quantities carrying anomalous dimensions remains an interesting open problem.

\subsection{Three--point functions}
The three--point functions for the superfields with arbitrary generalized conformal dimensions $\Delta_i$, $(i=1,2,3)$
\be
W(Z_1,Z_2, Z_3)= \langle \Phi(X_1, \theta_1) \Phi(X_2, \theta_2)   \Phi(X_3, \theta_3) \rangle \,,
\ee
may be computed in a way similar to the two--point functions using the superconformal Ward identities.
The invariance under $Q$--supersymmetry implies that they depend on the superinvariant intervals $Z_{ij}$,
\emph{i.e.}
\be
 \langle \Phi(X_1, \theta_1)  \Phi(X_2, \theta_2)   \Phi(X_3, \theta_3) \rangle= W(Z_{12}, Z_{23}, Z_{31})\,,
\ee
where
\be\label{differ}
Z_{ij}^{\mu \nu} = X_i^{\mu \nu} - X_j^{\mu \nu} - \frac{i}{2} (\theta_i^\mu \theta_j^\nu +\theta_i^\nu \theta_j^\mu)\,, \qquad i,j=1,2,3\,.
\ee
Invariance under $S$--supersymmetry then fixes  the form
of the function $W$ to be
\bea \label{3PTS}
	\begin{split}
&\langle \Phi(X_1, \theta_1) \Phi(X_2, \theta_2)   \Phi(X_3, \theta_3) \rangle = \\
&=c_3(\det Z_{12})^{-\frac{1}{2}(\Delta_1+\Delta_2 -\Delta_3)} (\det Z_{23})^{-\frac{1}{2}(\Delta_2+\Delta_3 -\Delta_1)}
 (\det Z_{31})^{-\frac{1}{2}(\Delta_3+\Delta_1 -\Delta_2)} \,.
 	\end{split}
\eea
Let us note that the three--point function is not annihilated by the operator entering the free equations of motion (\ref{DDPhi})
for generic values of the generalized conformal dimensions, including the case in which the values
of all the generalized conformal dimensions are canonical
 \bea \nonumber
&&(D^1_{\mu } D^1_{\nu }  - D^1_{\nu} D^1_{\mu } )
\langle \Phi(X_1, \theta_1), \Phi(X_2, \theta_2) , \Phi(X_2, \theta_2)    \rangle= \\ \nonumber
&&=c_3(D^1_{\mu } D^1_{\nu }  - D^1_{\nu} D^1_{\mu } ) \left (
({\rm det} |Z_{12}|)^{-\frac{1}{4}}({\rm det} |Z_{23}|)^{-\frac{1}{4}}   ({\rm det} |Z_{31}|)^{-\frac{1}{4}}  \right ) \neq 0 \,.
\eea
The component analysis of the superfield three--point correlation function (\ref{3PTS})
proceeds in the same way as in the ${\mathcal N}=1$, $D=3$ case of Section \ref{3pf-3d-n1}. The difference lies, however, in the presence of many more auxiliary fields.

Again, the three--point functions on the supergroup manifold $OSp(1|n)$ can be obtained via the Weyl rescaling
(\ref{W}), as in the case of the two--point functions, eq. \eqref{osp2p}.

\subsection{Four--point functions}

Finally, let us consider, first in flat hyper superspace, the correlation function of four real scalar superfields
with arbitrary generalized conformal dimensions,
 $\Delta_i$ (with $i=1,2,3,4$)
\be
W(Z_1,Z_2, Z_3)= \langle \Phi(X_1, \theta_1) \Phi(X_2, \theta_2)   \Phi(X_3, \theta_3)  \Phi(X_4, \theta_4)\rangle\,.
\ee
Invariance under $Q$--supersymmetry again implies that
the correlation function depends only on the superinvariant intervals
$Z_{ij}^{\mu \nu}$ \eqref{differ}. Following  the analogy with conventional conformal field theory we find
\begin{equation}\label{4PF}
W(X_1, X_2, X_3, X_4) = c_4\,\prod_{ij, i<j}\frac{1}{{(\det |Z_{ij}|)}^{k_{ij}}   }
{\tilde W} \left (  z,z' \right)\,,
\end{equation}
with $W$ being an arbitrary function of the  cross-ratios
\begin{equation}
	z= \det\left(\frac{|Z_{12}||Z_{34}|}{|Z_{13}||Z_{24}|}\right)~,\qquad z'= \det\left(\frac{|Z_{12}||Z_{34}|}{|Z_{23}||Z_{14}|}\right)\,,
\end{equation}
subject to the crossing symmetry constraints
\begin{equation}\label{crossingSym}
	 \tilde W(z,z')=\tilde W \left(\frac{1}{z},\frac{z'}{z}\right)=\tilde W \left(\frac{z}{z'},\frac{1}{z'}\right)\,.
\end{equation}
Furthermore, the $k_{ij}$'s are constrained by invariance of the four--point function under the $S$--supersymmetry to satisfy
\begin{equation}
\sum_{j\neq i} k_{ij} =  \Delta_i \,.
\end{equation}
Similarly to the case of two-- and three--point functions, the four--point function
of the scalar superfields on $OSp(1|n)$ can be obtained from (\ref{4PF}) via  the Weyl re--scaling (\ref{W}).

\section{Conclusion and outlook}

A detailed study of the  $OSp(1|2n)$--invariant generalized superconformal theories
is still an interesting open problem, which is
important for better understanding the
properties of conformally invariant higher-- spin field theories
(see \emph{e.g.} \cite{Gover:2008sw,Metsaev:2009ym,Metsaev:2013wza,Metsaev:2014iwa,Costa:2011mg,Maldacena:2011jn,Stanev:2012nq,Giombi:2013yva,Nutma:2014pua} for  recent progress in studying conformal higher-- spin fields).
 Our results are a further step in this direction. Following the program outlined in \cite{Florakis:2014kfa},
 we have extended the results on the structure of
$Sp(2n)$--invariant field equations to supersymmetric higher-- spin systems. We constructed generalized superconformal transformations relating the field equation on flat hyper--superspace and on
$OSp(1|n)$ supergroup manifolds, which correspond to a generalization of supersymmetric AdS spaces.
We computed the two--, three-- and four--point functions
of real hyper--superfields
both on flat and on $OSp(1|n)$
supergroup manifolds
and, as a simple  illustration of our approach, applied this
technique to the example of ${\mathcal N}=1$, $D=3$
superconformal theory of scalar superfields.

It is important
to further study possible interactions (which might be associated with non--trivial three-- and four--point correlation functions)
in this type of models. Since a Lagrangian description of $OSp(1|2n)$ invariant field equations is still not known even in the free case, one can approach the problem using non--Lagrangian methods similar to those in Conformal Field Theories (see for example \cite{Petkou:1994ad}).
Following these methods one can try to introduce $OSp(1|2n)$ invariant vertexes  and  compute  explicit expressions  for
anomalous dimensions for generalized conformal weights.  Recall that according to the results of Section \ref{s5}
the Ward identities for three-- and four--point functions do not necessarily require the values of the generalized conformal weights to be canonical, therefore one may expect interesting outcomes of this study.

The question of the existence of anomalous values for generalized conformal dimensions can be related to the question
of a possible breaking of $OSp(1|2n)$ symmetry down to a corresponding $AdS_D$ (super)symmetry.
In this respect one can also note that the hyperspace formulation considered in this paper does not involve higher--spin gauge field potentials, but only their field strengths. So far higher--spin potentials have been introduced only in an unfolded extension of the hyperspace formulation of $D=4$ higher--spin fields in such a way that the resulting equations are invariant under $SU(2,2)$  and $O(3,3)$  subgroups of the original $Sp(8)$ symmetry, motivating to speculate on their origin due to a mechanism of spontaneous breaking of higher--spin and $Sp(8)$ symmetries \cite{Vasiliev:2007yc}. Further study in this direction may help in searching for interacting systems of fields on hyper-(super)spaces and their possible connection to Vasiliev's interacting higher--spin gauge theories.

It would be also of interest to consider in detail the implication of our results in the framework of
higher--spin AdS/CFT correspondence. The origin of higher--spin holographic duality can be
traced back \cite{Vasiliev:2001zy} to the work of Flato and Fronsdal \cite{Flato:1978qz} who showed that the tensor
product of single-particle states of a $3D$ massless conformal scalar and spinor fields (singletons) produces the tower of
all single--particle representations of $4D$ massless fields whose spectrum matches
that of $4D$ higher--spin gauge theories.
The hyperspace formulation provides an explicit field theoretical realization
of the Flato--Fronsdal theorem in which higher--spin fields are also ``packed" in a single scalar and spinor fields, though propagating in hyperspace.
The relevance of the hyperspace formulation to holography has been pointed out in \cite{Vasiliev:2001zy,Vasiliev:2012vf}. In this interpretation, holographically dual theories share the same
unfolded formulation in extended spaces which contains  twistor--like (or oscillator) variables and each of these theories corresponds to a different reduction, or  ``visualization", of the same ``master" theory.
For instance, the higher--spin field equations in either ordinary space--time or hyperspace can be obtained
from the same set of unfolded equations \cite{Vasiliev:2001zy,Didenko:2003aa,Plyushchay:2003gv,Plyushchay:2003tj}.
Depending on the number of twistorial coordinates of the unfolded formulation, one can obtain hyperfields of
different ranks which can be fundamental fields, bi--fundamental fields (currents) etc. \cite{Gelfond:2003vh}.
A connection between these fields in different dimensions can be established
via embedding of lower--dimensional hyperspaces into higher-- dimensional ones \cite{Gelfond:2010pm}.
Thus, one can conclude that the hyperspace formulation provides an  extra and potentially powerful tool
for  studying higher--spin AdS/CFT correspondence.

 A detailed study of the higher--spin content of field equations on higher--dimensional curved hyper--superspaces,
as well as their underlying higher--spin superalgebras containing $OSp(1|n)$, is yet another interesting issue.
 We hope to address these problems in future work.

\subsection*{\bf Acknowledgments}
We are grateful to I.Bandos,  N. Berkovits,   S. Kuzenko,  I. Samsonov,  M. Vasiliev and P. West for fruitful discussions.
The work of D.S. was partially supported by the Padova University Project CPDA119349, the INFN Special Initiative ST\&FI and by the Russian Science Foundation grant 14-42-00047 in association with Lebedev Physical Institute. D.S. would also like to acknowledge the warm hospitality extended to him at the Faculty of Education, Science, Technology and Mathematics, University of
Canberra, during an intermediate stage of this work.
M.T. would like to thank the Department of Physics, the University of Auckland, where part of this work has
been performed, for its kind hospitality. The work of M.T.  has been supported in part by an  Australian Research Council  grant DP120101340. M.T. would also like to acknowledge grant   31/89 of the Shota Rustaveli National Science Foundation.

\if{}
\bibliographystyle{utphys}
\bibliography{references}
\end{document}
\fi

\providecommand{\href}[2]{#2}\begingroup\raggedright\endgroup

\end{document}